\documentclass[11pt]{article}

\usepackage{graphicx}
\usepackage{bm}

\begin{document}


\title{Tunneling spectroscopy of topological superconductors}

\author{
\textsc{Satoshi. Kashiwaya},$^{1}$ 
\textsc{Hiromi Kashiwaya},$^{1}$ 
\textsc{Kohta Saito},$^{1}$\\
\textsc{Yasunori Mawatari},$^{1}$ 
and \textsc{Yukio Tanaka},$^{2}$}
\maketitle


$^1$National Institute of Advanced Industrial Science and Technology (AIST), Tsukuba 305-8568, Japan\\
$^2$Department of Applied Physics, Nagoya University, Nagoya 464-8603, Japan\\


\date{\today}

\begin{abstract}
Tunneling conductance spectra of normal metal/insulator/superconductor ($N/I/S$) junctions are calculated to determine the potential of tunneling spectroscopy in investigations of topological superconductivity.
Peculiar feature of topological superconductors is the formation of gapless edge states in them. 
Since the conductance of $N/I/S$ junctions is sensitive to the formation of these edge states, topological superconductivity can be identified through edge-state detection. 
Herein, the effects of Fermi surface anisotropy and an applied magnetic field on the conductance spectra are analyzed to gather indications that can help to identify the topological nature of actual materials.
\end{abstract}

\section{Introduction}
\label{1}
Tunneling spectroscopy has been accepted as one of the highest energy resolution probes for electronic states. 
Various novel phenomena relevant to superconductivity have been revealed by tunneling spectroscopy of normal metal/insulator/superconductor ($N/I/S$) junctions. 
Recently, there is an increasing interest in the properties of topological superconductors (TSCs), whose superconducting gap function in momentum space is topologically non-trivial [1-6]. 
Although novel properties of the TSCs, such as Majorana fermions at the edges and the vortex cores, have been predicted in a number of theories [7-15], the focus of most experiments remains at the level of exploring the exisitence of TSCs. 
The identification of the TSCs based on conventional experimental probes is difficult because the bulk densities of states (DOS) of most of the TSCs are similar to those of conventional superconductors. 
On the other hand, a peculiar feature of topological materials, the so-called "bulk-edge correspondence," is known to induce gapless edge states formed at the edges and defects at which the translational symmetry is broken [16-18]. 
Therefore, the detection of these gapless states inside the superconducting gap is the manifestation of topological superconductivity in real materials.
\par
On the other hand, unconventional superconductivity is characterized by the anisotropic gap function, whose amplitude and phase depend on the wave vector.
This anisotropic gap function is known to induce Andreev bound states (ABSs) at the surfaces and interfaces where the effective pair potential for the quasiparticles changes through reflection.
The formation of ABSs has been predicted for the surface states of $^3$He [19, 20] and described for the surface states of d$_{x^2-y^2}$-wave superconductors [21-27].
Recently the zero-energy ABS has been reinterpreted as a topological edge state [18, 28-32]; therefore, the high sensitivity of tunneling spectroscopy to surface states can make a significant contribution to our understanding of TSCs based on the observation of the edge states.
Experimentally, the existence of the ABS has already been verified by the zero-bias peaks observed in the conductance spectra of $N/I/S$ junctions of various high-$T_c$ cuprates [22,25].
Furthermore, the realization of the topological superconductivity and superfluidity has been suggested for Sr$_2$RuO$_4$, Cu-doped Bi$_2$Se$_3$, PdBi, and $^3$He [33-36].
\par
Here, we systematically analyze the conductance spectra of TSCs with various types of gap functions by comparing them with those of non-topological superconductors.
We also examine the correspondence between topological superconductivity and the conductance peak shape with taking account of the sensitivity to the orientation of the tunneling junction and the effects of anisotropy. 
Furthermore, we discuss how to detect broken time-reversal symmetry (TRS) based on the magnetic field response of $N/I/S$.
\par
\section{Conductance spectra of TSCs}
\label{sec2}
Herein, we consider an $N/I/S$ junction with a flat interface perpendicular to the $x$-axis, as shown in Fig. 1(a). 
Although the superconductor is assumed to be quasi-two-dimensional with an isotropic Fermi surface in the $x$-$y$ plane, for simplicity, all the results can be directly extended to the three-dimensional case. 
Two-dimensionality has an advantage in that the ABSs do not appear on the $z$-plane, so the bulk DOS can be probed using a $z$-axis tunneling junction without the influence of the edge states [33].
\par
The gap function of the spin-singlet superconductor is represented by $\Delta$($\mathbf{k}$), where $\mathbf{k}$ is the wave vector of the quasiparticle in the $xy$-plane, and of the triplet $\Delta_{\mathbf{k}}=[\mathbf{d}_k,\mathbf{\sigma}]i \mathbf{\sigma_y}$, where $\mathbf{\sigma}$ and $\mathbf{d}_k$ are Pauli matrices and the $\mathbf{d}$-vector, respectively. 
The angle-resolved conductance spectrum $\sigma$($\theta$, $eV$) for the quasiparticle injected from the left, with the incident angle $\theta$ and energy $eV$ to the interface, is calculated through the extension of the BTK model [37]. 
The total conductance $\sigma$($eV$) is given by integrating over all injection angles. 
Details of the formulation for singlet cases are described in [23, 25] and for triplet cases in [18, 38-41]. 
The conductance for a given $\theta$ contains two distinct gap functions $\Delta_{+}$ and $\Delta_{-}$, which correspond to the effective pair potentials for the transmitted electron-like and hole-like quasiparticles, respectively. 
The barrier potential of the insulator and the barrier parameter $Z$ are represented by $H\delta(x)$ and $Z=mH/(\hbar^2k_F)$, respectively, where $m$, $H$, and $k_F$ are the electron mass, amplitude of the barrier potential, and Fermi wavelength, respectively.
\par
We consider eight types of pairing symmetries, as listed in Table I, by mainly taking into account of the possible candicates for Sr$_2$RuO$_4$.
Figures 2 show the calculated conductance spectra $\sigma(\theta, eV)$ for these gap functions with $Z$=4 plotted as a function of the injection angle $\theta$. 
The red color regions inside the gap amplitude indicate high conductance regions resulting from the formation of the ABS at the edge of the superconductors. 
The zero-energy ABS does not exist for all injection angles for non-topological superconductors, whereas the zero-energy ABS does emerge for topological superconductors [4,5,6,18,42]. 
It is well known that the zero energy ABS is formed when the phase difference between $\Delta_{+}$ and $\Delta_{-}$ is $\pi$. 
In the case of a $d_{xy}$-wave, the phase difference is $\pi$ for all $\theta$, thus the flat zero-energy ABS are formed for all $\theta$ [30,43].
The flat ABS is easily distorted due to the presence of a subdominant $s$-wave component as shown by $d_{xy}$+$is$-wave case[44].
On the other hand, in the cases of chiral and helical superconductors, the phase difference strongly depends on $\theta$, and the ABS forms a diagonal or cross-shaped distribution with an almost linear gradient near the Fermi level [4,18,28,29,39-41]. 
The asymmetric ABS with respect to the inversion of $\theta$, obtained for the chiral $d$-wave, the chiral $p$-wave, the $d_{xy}$+$is$-wave cases, is the manifestation of the broken TRS. 
As a result of the asymmetry, net current flow at the edge is expected to appear in these superconductors [45,46].
\par
Figure 3 shows the corresponding $\theta$-integrated conductance spectra $\sigma$($eV$) for 8 kinds of pairing symmetries. 
For all topological cases, the conductance spectra exhibit the zero-bias peak, although the sharpness of the peak depends on the ABS distribution. 
However, it should be noted that the appearance of the zero-bias peak in a tunneling experiment does not always certify the presence of the topological superconductivity, as discussed in the next section. 
Therefore, the clarifying of the $\theta$-resolved nature of the ABS is quite important in addition to tunneling spectroscopy. 
It becomes possible if the energy resolution of the angle-resolved photoemission spectroscopy (ARPES) improves sufficiently to detect the ABS with $\mathbf{k}$-resolution.
\section{Effects of anisotropy}
\label{3}
In real experimental situations, there always exist deviations from the ideal isotropic model discussed above. 
For example, specific heat measurements have shown the presence of the large anisotropy in the gap amplitude of Sr$_2$RuO$_4$ [47,48]. 
Although the conductance spectrum of the chiral $p$-wave is insensitive to the junction orientation if the pair amplitude is isotropic, it becomes highly sensitive to the orientation if anisotropy exists [39, 40]. 
A wide variety of conductance spectra has actually been reported in the tunneling spectroscopy experiments of Sr$_2$RuO$_4$ [33]. 
Here, we investigate the effects of anisotropy and the misorientation on the conductance spectra to examine the origin of sample dependence.
\par
In the following, we analyzed conductance spectra based on a phenomenological model of the anisotropic Fermi surface. 
We introduce the anisotropy by replacing $\Delta_0$ by $\Delta_0(1+C\cos4\theta)$, (where $C$ is a parameter describing the degree of anisotropy; see Fig. 1(b)) referring to the results of a specific heat measurement on Sr$_2$RuO$_4$. 
The misorientation angle $\alpha$ is defined by the angle between the $a$-axis of the crystal and the normal interface, and the $c$-axis is assumed to be parallel to the $z$-axis. 
Figure 4(a) shows the angle-resolved conductance spectra with anisotropic pair amplitude ($C$=0.3) and without misorientation; Figure 4(b) shows the $\theta$-integrated conductance spectra ($C$=0, 0.1, 0.2, 0.3). 
It is evident that the $\theta$-dependence of ABS is modified due to the anisotropy and the zero-bias peak changes to the dip, even though the zero-energy ABS persists. 
Figure 5(a) shows the angle-resolved conductance spectra with anisotropy and with misorientation ($\alpha$=0.125$\pi$). 
Although the misorientation has no effect on the conductance spectra for the isotropic case, it has a serious effect here, as shown by the appearance of multiple ripples in the $\theta$-dependence of the ABS. 
As a result, the $\theta$-integrated conductance spectra exhibit complex peaks and dips, as described in Fig. 5(b). 
On the other hand, the zero-bias conductance (ZBC) is stable against these purturvations reflecting the fact that the single zero-energy ABS is topologically protected.
\par
The sensitivity of the chiral $p$-wave is evidently responsible for the linearly dispersive ABS [49,50].
Such a variety of spectra does not appear for the $d_{xy}$-wave case, because the phase difference between $\Delta_{+}$ and $\Delta_{-}$ is always fixed to 0 or $\pi$, whereas the phase difference between $\Delta_{+}$ and $\Delta_{-}$ can take arbitrary value in the case of the chiral $p$-wave.
Based on above considerations, the sample dependence of the conductance spectra detected in experiments is suggested to stem from the anisotropy and the uncontrolled misalignment at the junction interface. 
It is also shown that the lack of the zero-bias conductance peak in experiments does not always mean topologically trivial superconductivity.
\section{Magnetic field response and broken TRS}
\label{4}
There are several candidates for the pairing symmetry of Sr$_2$RuO$_4$ based on a number of experimental works on this material. 
The chiral $p$-wave ($L_z$=1, -1, $L_z$ is orbital angular momentum) and the helical $p$-wave are the most plausible candidates among them [48].
Unfortunately, tunneling spectroscopy cannot discriminate between these three types as shown in Fig. 3(b). 
Here, we present a useful way to experimentally discriminate these pairing symmetries based on the magnetic field response of the conductance spectra [51,52].
\par
One of the peculiarities of the chiral superconductor is the presence of a spontaneous edge current arising from the orbital moment of the Cooper pairs. 
When an external magnetic field $H$ is applied along the $c$-axis, the Messiner effect also induces the finite shielding current at the edge of the superconductor. 
In the case of the chiral $p$-wave, the Messiner current enhances or suppresses the spontaneous edge current depending on the relative direction between the applied magnetic field and the chirality. 
On the other hand, such asymmetric responses cannot be expected for the helical $p$-wave that preserves the TRS. 
The net current flow at the edge can be detected by the conductance spectra through the Doppler shift of the quasiparticle energy [45]. 
Similar effects of the Doppler shift have been discussed in the context of the spontaneous symmetry-breaking in high-$T_c$ cuprates: the origin of the splittings of the ZBCP detected in tunneling experiments has been attributed to the transition from the $d_{xy}$-wave to the $d_{xy}$+$is$-wave [45,53,54]. 
Although the broken TRS is common both to the $d_{xy}$+$is$-wave and the chiral $p$-wave, $\theta$-dependences of the ABS are quite different in that no zero-energy ABS exists for the $d_{xy}$+$is$-wave as shown in Fig. 2(d) and (f). 
In the cases of $d_{xy}$+$is$-wave, the $\theta$ dependence is rather weak compared to the chiral $p$-wave; thus, the conductance peak is sufficiently sharp to detect the splitting. 
On the other hand, in the case of the chiral $p$-wave, the effect of the Doppler shift can be deduced only by the asymmetric response due to the widely distributed ABS.
\par
Figure 6 shows the angle-resolved conductance spectra of the chiral $p$-wave when the magnetic field $H$ (0.2 $H_c$, where $H_c$ is the critical magnetic field given by $\Delta_0/(e\lambda V_F)$, $\lambda$ and $V_F$ are the penetration depth and the Fermi velocity, respectively) is parallel and anti-parallel to $L_z$, respectively. 
When the magnetic field is applied, $\theta$-dependence of the ABS is distorted to increase or decrease the gradient; consequently, the conductance peak height is suppressed or enhanced as shown in Fig. 6(a) and (b). 
Figure 6(c) shows the dependence of the ZBC to the applied field. 
It is evident that the three types of symmetries can be easily distinguished by the magnetic field response.
\par
We comment on whether the broken TRS in Sr$_2$RuO$_4$ can be detected by tunneling spectroscopy. 
In chiral $p$-wave superconductors, the domain structure of $L_z$=1 and $L_z$=-1 regions is expected to emerge similar to the magnetic domains in ferromagnetic materials. 
When the domain size was smaller than the junction dimension, we could only observe spatially averaged conductance over multiple domains. 
Therefore, the junction dimension should be designed as small as possible to detect single domain features.
\section{Conclusion}
\label{5}
The conductance spectra of normal metal/insulator/superconductor junctions are calculated for various types of pairing symmetries that are possible candidates for Sr$_2$RuO$_4$. 
The formation of the zero-energy Andreev bound state is the manifestation of topological superconductivity, whereas the presence or the absence of the zero-bias conductance peak does not directly correspond to topological or non-topological superconductivity. 
This is because the peak shape of the conductance spectrum is sensitive to the details of the junction, such as the anisotropy of the pair amplitude and the misorientation, particularly when the ABS has a strong $\theta$-dependence. 
The broken time-reversal symmetry can be tested by the magnetic field response of the conductance peak through the Doppler shift of quasiparticle energy. 
We expect the results presented in this paper will help in identifying the topological nature of novel superconductors. 
\section{Acknowledgements}
\label{6}
This work was financially supported by a Grant-in-Aid for Scientific Research on Innovative Areas "Topological Quantum Phenomena" (No.~22103002) from MEXT, and by a Grants-in-Aid for Scientific Research (No.~25286046, ~25820150) from JSPS, Japan.

\bibliographystyle{model1a-num-names}
\bibliography{<your-bib-database>}



\newpage
\includegraphics[width=1\linewidth]{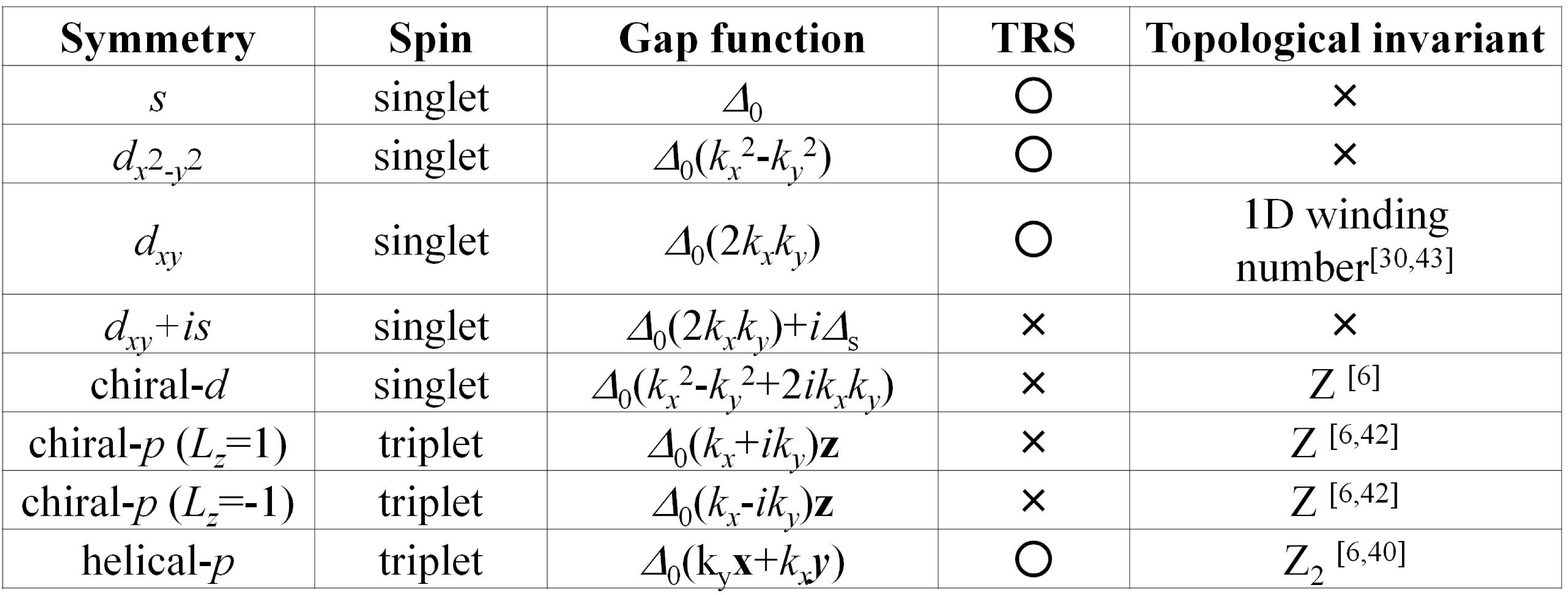}
Table 1. Eight pairing symmetries and their gap functions, spin states, time reversal symmetry (TRS), and topological invariants used in the present calculation are summarized.
$L_z$ is orbital angular momentum of a Cooper pair, and 1D means one-dimensional.
\newpage

\begin{figure}[t]
\includegraphics[width=1\linewidth]{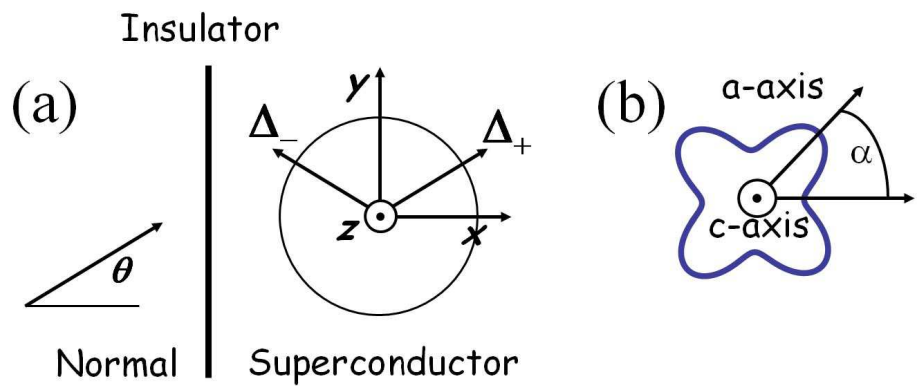}
\caption{\label{fig1}
(Color online)
(a) Schematic of a normal metal/insulator/superconductor junction and the quasiparticle injection from the normal metal. 
(b) Schematic of a anisotropy of the gap amplitude.
}
\end{figure}
%
\begin{figure}[t]
\includegraphics[width=1\linewidth]{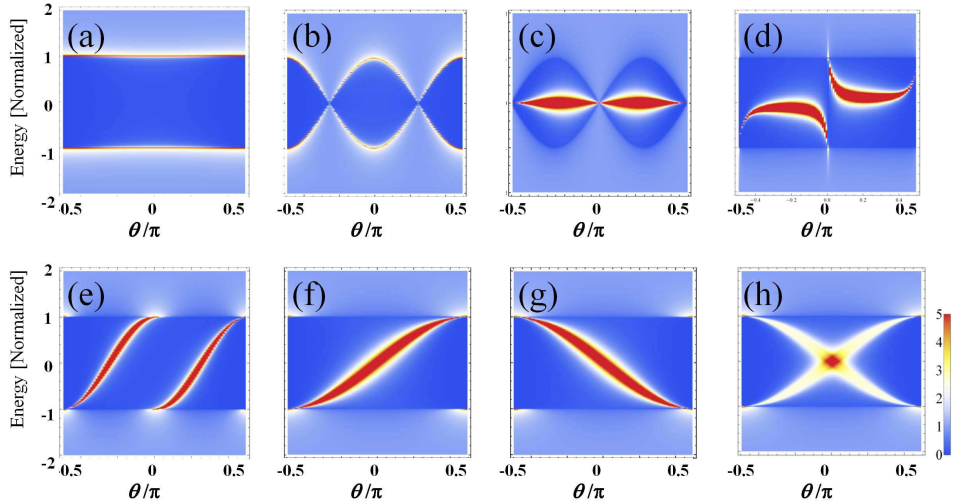}
\caption{\label{fig2}
(Color online)
Angle resolved conductance spectra for the eight pairing symmetries with the barrier parameter Z=4,
(a) $s$-wave, (b) $d_{x^2-y^2}$-wave, (c) $d_{xy}$-wave, (d) $d_{xy}$+$is$-wave, (e) chiral $d$-wave, (f) chiral $p$-wave ($L_z$=1), (g) chiral $p$-wave ($L_z$=-1), (h) helical $p$-wave.
The color represents the magnitude of conductance of the junction, the vertical axis is the quasiparticle energy normalized by $\Delta_0$, and the horizontal axis is the injection angle $\theta$. 
Since the horizontal axis can be converted to a wave vector $k_y$(=$k_F\sin\theta$), we can easily recognize the dispersion of the ABS from these figures.
Red regions inside the gap correspond to the conductance peak originating from the ABS formed at the edge. When the ABS sits on the zero-energy level at a certain angle, the superconductor becomes topologically non-trivial.
The amplitude of $\Delta_S$ is assumed to 0.2$\Delta_0$.
}
\end{figure}
\begin{figure}[t]
\includegraphics[width=1\linewidth]{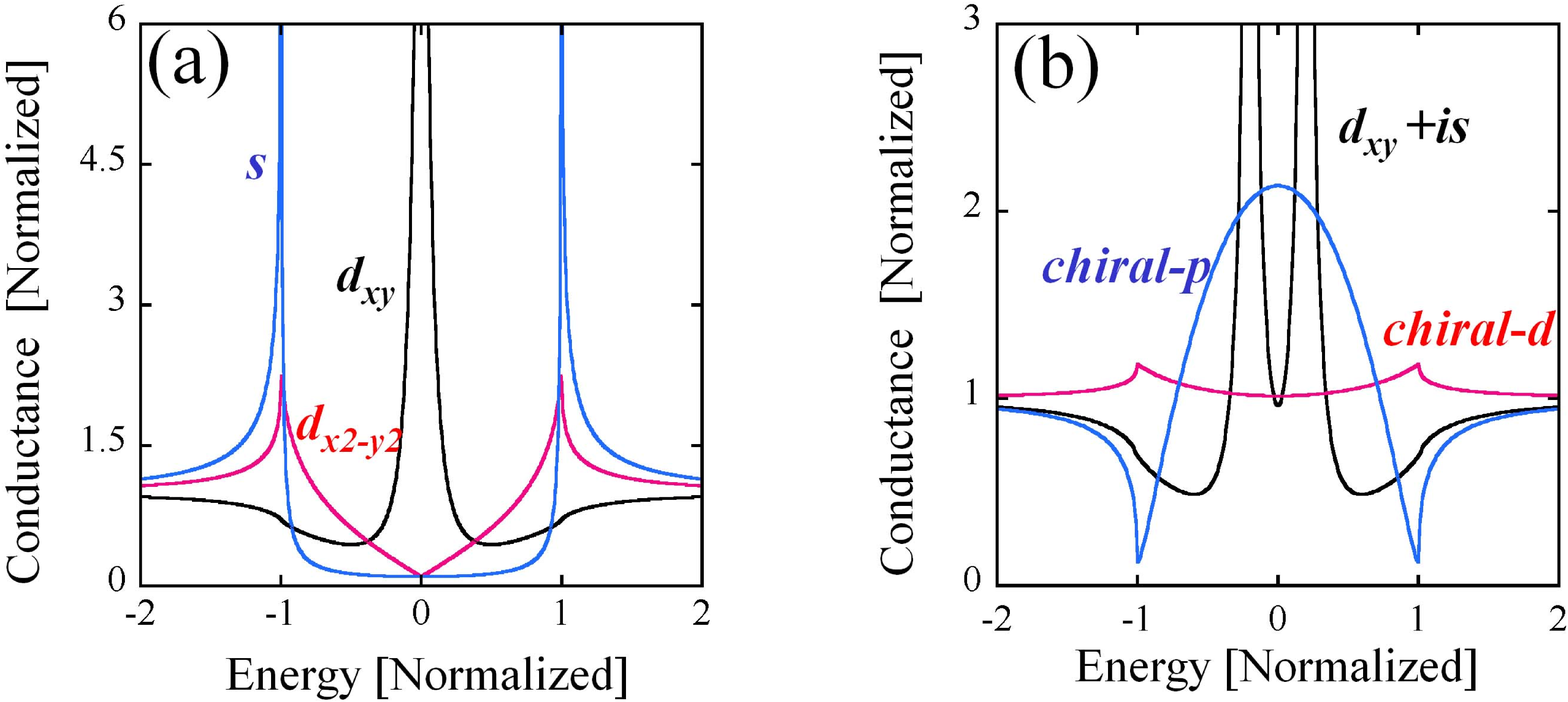}
\caption{\label{fig3}
(Color online)
Conductance spectra for the eight pairing symmetries.
The horizontal and vertical axes are normalized by $\Delta_0$ and normal conductance, respectively.
(a) The conductance spectra for (i) $s$-wave, (ii) $d_{x^2-y^2}$-wave, (iii) $d_{xy}$-wave,
(b) (i) $d_{xy}$+$is$-wave, (ii) Chiral $d$-wave, (iii) chiral $p$-wave ($L_z$=1), (iv) chiral $p$-wave ($L_z$=-1), (v) helical $p$-wave.
The conductance spectra for the latter three types fall onto the same plot.
}
\end{figure}
\begin{figure}[t]
\includegraphics[width=1\linewidth]{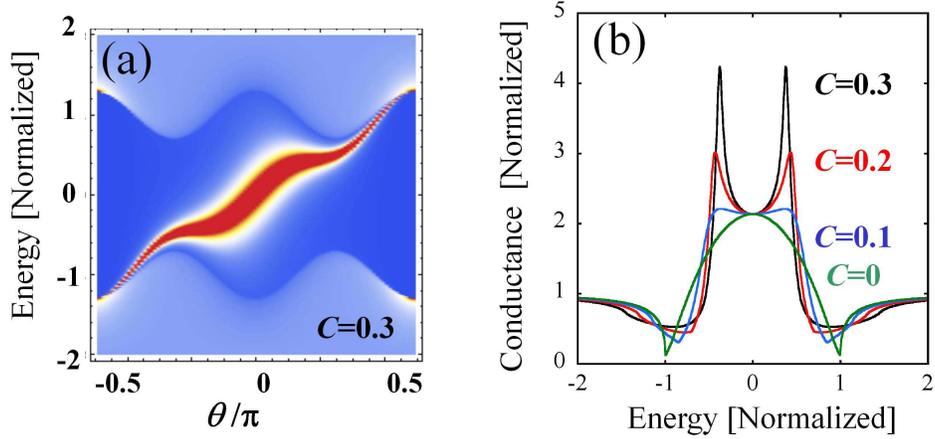}
\caption{\label{fig4}
(Color online)
The angle-resolved conductance spectra of the chiral $p$-wave with anisotropic pair amplitude ($C$=0.3) and without misorientation. 
(b) The $\theta$-integrated conductance spectra for $C$=0, 0.1, 0.2, and 0.3 without misorientation. 
The horizontal and vertical axes are normalized by $\Delta_0$ and normal conductance, respectively.
Although the zero-energy ABS exists robustly, the conductance peak shape changes to dips by increasing the anisotropy.
}
\end{figure}
\begin{figure}[t]
\includegraphics[width=1\linewidth]{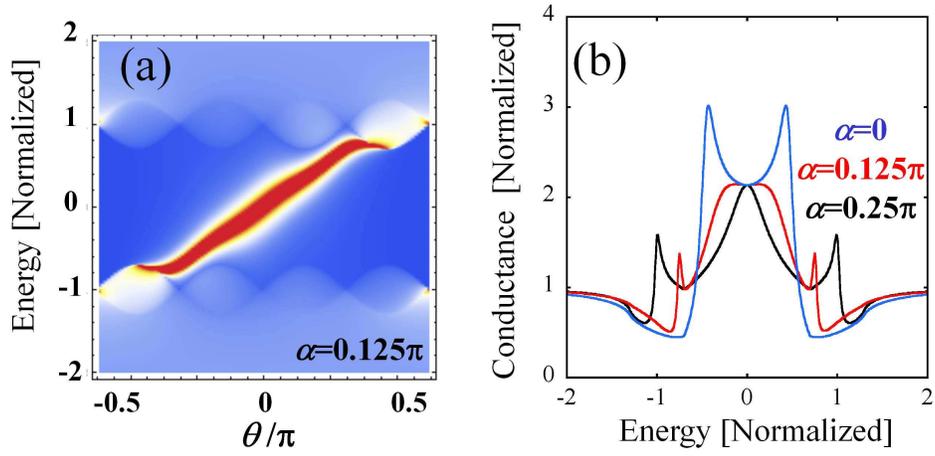}
\caption{\label{fig5}
(Color online)
The angle-resolved conductance spectra of the chiral $p$-wave with anisotropic pair amplitude ($C$=0.2) and with misorientation ($\alpha$=0.125$\pi$). 
(b) The $\theta$-integrated conductance spectra for $C$=0.2 and $\alpha$=0, $\alpha$=0.125$\pi$, and $\alpha$=0.25$\pi$. 
The horizontal and vertical axes are normalized by $\Delta_0$ and normal conductance, respectively.
The conductance spectra show complex peak and dip structures.
}
\end{figure}
\begin{figure}[t]
\includegraphics[width=1\linewidth]{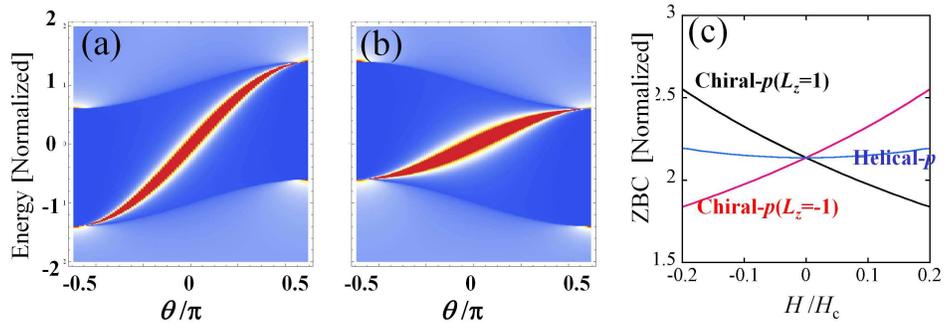}
\caption{\label{fig6}
(Color online)
(a, b) The angle-resolved conductance spectra of the chiral $p$-wave ($L_z$=1), the chiral $p$-wave ($L_z$=-1) with the magnetic field ($H$=0.2$H_c$) parallel to the $c$-axis being applied. 
(c) The magnetic field responses of the zero-bias conductance (ZBC) for the chiral $p$-wave ($L_z$=1), the chiral $p$-wave ($L_z$=-1), and the helical $p$-wave.
The magnetic field is normalized by $H_c$.
The broken TRS can be identified experimentally by the asymmetric response of the ZBC to the applied field.
}
\end{figure}

\end{document}